\documentclass[aps, preprint, showkeys]{revtex4-1}
\usepackage{amsmath}
\usepackage{amsfonts}
\usepackage{graphicx}
\usepackage{amssymb}
\usepackage{graphics}
\usepackage{float}
\usepackage[font=scriptsize]{caption}

\begin{document}
\title{Optimal Local Transformations of Flip and Exchange Symmetric Entangled States}

\author{G. Karpat}
\email{e-mail: gkarpat@sabanciuniv.edu}
\affiliation{Faculty of Engineering and Natural Sciences, Sabanci University, Tuzla, Istanbul 34956, Turkey}

\author{Z. Gedik}
\affiliation{Faculty of Engineering and Natural Sciences, Sabanci University, Tuzla, Istanbul 34956, Turkey}

\date{\today}

\begin{abstract}
Local quantum operations relating multiqubit flip (0-1) and exchange symmetric (FES) states, with the maximum possible probability of success, have been determined by assuming that the states are converted via one-shot FES transformations. It has been shown that certain entangled states are more robust than others, in the sense that the optimum probability of converting these robust states to the states lying in the close neighborhood of separable ones vanish under local FES operations.
\end{abstract}


\maketitle

\section{Introduction}

Quantum entanglement is considered as a key resource in quantum computation, quantum cryptography and quantum information processing \cite{1}. One of the most important problems in quantum information theory is whether it is possible
to convert an entangled state, that is being shared between two spatially separated parties, to another by applying only local quantum operations on each subsystem. If two parties are also allowed to communicate classically this entanglement manipulation scheme is called local operations and classical communication (LOCC) \cite{2,3}. Two pure states can be obtained with certainty from each other by means of LOCC if and only if they are related by local unitary operations. The condition of certainty can be removed to allow probabilistic conversion of states through stochastic local operations and classical communication (SLOCC) \cite{4}. This coarse-graining simplifies the equivalence classes labeled by continuous parameters in case of the local unitary operations. With the consideration of SLOCC, two states are said to have same kind of entanglement if an invertible local operation (ILO) relating them exists \cite{5}.

In case of pure two-qubit states, there is a single equivalence class, i.e., all entangled states are equivalent to EPR state $(1/\sqrt{2})(|00\rangle+|11\rangle)$ under SLOCC \cite{5,6}. For three qubits, it was shown that entangled states can be converted either to the GHZ state $(1/\sqrt{2})(|000\rangle+|111\rangle)$, or to the W state $(1/\sqrt{3})(|001\rangle+|010\rangle+|100\rangle)$ by the application of SLOCC operations \cite{5}. In other words, three qubits can be entangled in two inequivalent ways. Although two and three qubit entanglement is relatively well understood, classification problem becomes complicated for multiqubit systems starting from four qubits since continuous parameters are needed to label the equivalence classes \cite{5}. The classification of four-qubit entangled states has also been done yet there are different (but complementary) points of view on the criteria of classification \cite{7,8,9,10}. There are no complete classifications for five or more qubits. However, for $n$-qubit exchange symmetric states, SLOCC classification has been obtained with the help of Majorana representation of symmetric states \cite{11}.

Despite the fact that finding an ILO which relates two states with some non-zero probability is sufficient to show the equivalence of these two states under SLOCC, the success probabilities of transformations have fundamental operational importance in quantum information processes. For pure bipartite states, the transformations relating two states of the same class with the greatest probability of success have been found both in the cases of allowing and forbidding classical communication between the parties \cite{6,12,13,14}. While the complete solution of the problem is not known for the states involving three or more qubits, there are several works in the literature providing partial solutions \cite{15,16,17,18,19,20}.

In this work, flip and exchange symmetric (FES) states, which are invariant when two qubits are interchanged or when all 0s (1s) are changed to 1s (0s), are considered. It has been recently shown that  many-qubit FES states constitute a set of curves in the Hilbert space and equivalence classes of these states under ILOs can be determined in a systematic way for an arbitrary number of qubits \cite{21}. The main purpose of the present work is to investigate the optimal local FES transformations relating two multiqubit FES states assuming that spatially separated parties are only allowed to apply one-shot local operations on their subsystems, i.e., they are not allowed to make use of classical communication. Although the coordination of local operations by the assistance of classical communication has been shown to enhance the power of transformations in certain cases \cite{6}, it has also been noted that classical communication is expensive in some situations \cite{22}.

\section{Constraints on the Elements of a Quantum Operation}

Necessary and sufficient conditions for the entries of a two by two matrix in order for the matrix to be an element of a single qubit quantum operation can be obtained directly from the probability-sum condition of quantum measurements. Consider two operation elements $M_{1}$ and $M_{2}$, which are $2\times2$ matrices, and the quantum operation $\rho\rightarrow\Phi(\rho)=M_{1} \rho M_{1}^{\dag}+M_{2} \rho M_{2}^{\dag}$ performed on a single qubit. The only requirement on the operation elements to come from a positive operator-valued measure (POVM) is the normalization condition that $M_{1}^{\dag}M_{1} + M_{2}^{\dag}M_{2}=I$, where $I$ denotes the $2\times2$ identity matrix and $M_{1}$ and $M_{2}$ are defined as
\begin{eqnarray}
M_{1}=
\left(\begin{array}{cc}
a_{1} & a_{2} \\
a_{3} & a_{4} \\
\end{array}\right)
\quad
\textup{and}
\quad
M_{2}=
\left(\begin{array}{cc}
a_{5} & a_{6} \\
a_{7} & a_{8} \\
\end{array}\right).
\end{eqnarray}
For diagonal elements, the normalization condition requires that
\begin{eqnarray}
|a_{1}|^{2}+|a_{3}|^{2}\leq 1
\quad
\textup{and}
\quad
|a_{2}|^{2}+|a_{4}|^{2}\leq 1.
\end{eqnarray}
Let us introduce the four dimensional state vectors
\begin{eqnarray}
|v_{o}\rangle =
\left(\begin{array}{cc}
|v_{ou}\rangle \\
|v_{od}\rangle \\
\end{array}\right) =
\left(\begin{array}{cccc}
a_{1} \\
a_{3} \\
a_{5} \\
a_{7} \\
\end{array}\right)
\quad
\textup{and}
\quad
|v_{e}\rangle =
\left(\begin{array}{cc}
|v_{eu}\rangle \\
|v_{ed}\rangle \\
\end{array}\right) =
\left(\begin{array}{cccc}
a_{2} \\
a_{4} \\
a_{6} \\
a_{8} \\
\end{array}\right)
\end{eqnarray}
with
\begin{eqnarray}
|v_{ou}\rangle=(a_{1}, a_{3})^{T},  \quad |v_{od}\rangle=(a_{5}, a_{7})^{T},\nonumber \\
|v_{eu}\rangle=(a_{2}, a_{4})^{T},  \quad |v_{ed}\rangle=(a_{6}, a_{8})^{T}.
\end{eqnarray}
So that one have
\begin{eqnarray}
M_{1}^{\dag}M_{1} + M_{2}^{\dag}M_{2} =
\left(\begin{array}{cc}
\langle v_{o} | v_{o} \rangle & \langle v_{o} | v_{e} \rangle \\
\langle v_{e} | v_{o} \rangle & \langle v_{e} | v_{e} \rangle \\
\end{array}\right)
=I.
\end{eqnarray}
Therefore, $\langle v_{ou} | v_{ou} \rangle+\langle v_{od} | v_{od} \rangle=\langle v_{eu} | v_{eu} \rangle+\langle v_{ed} | v_{ed} \rangle=1$ and $\langle v_{eu} | v_{ou} \rangle=-\langle v_{ed} | v_{od} \rangle$. The Schwarz inequality $|\langle v_{ed} | v_{od} \rangle|^{2}  \leq \langle v_{od} | v_{od} \rangle \langle v_{ed} | v_{ed} \rangle$ implies that
\begin{eqnarray}
|\langle v_{eu} | v_{ou} \rangle|^{2}  \leq (1-\langle v_{ou} | v_{ou} \rangle)(1- \langle v_{eu} | v_{eu} \rangle).
\end{eqnarray}
Writing the vectors in terms of $a_{i}$'s we obtain
\begin{eqnarray}
|a_{1}|^{2}+|a_{2}|^{2}+|a_{3}|^{2}+|a_{4}|^{2} \leq 1 + |\Delta|^{2}
\end{eqnarray}
where $\Delta=a_{1}a_{4}-a_{2}a_{3}$ denotes the determinant of the operation element $M_{1}$. On the other hand, if $a_{1},a_{2},a_{3},a_{4}$ are given with $|a_{5}|^{2}+|a_{7}|^{2}=1-|a_{1}|^{2}+|a_{3}|^{2}$ and $|a_{6}|^{2}+|a_{8}|^{2}=1-|a_{2}|^{2}+|a_{4}|^{2}$, Eq. (7) will ensure that $\langle v_{eu} | v_{ou} \rangle=-\langle v_{ed} | v_{od} \rangle$. Hence, Eq. (2) together with Eq. (7) give the necessary and sufficient conditions for a $2\times2$ matrix to be a valid operation element. Moreover, one can show that the inequalities given by Eq. (2) are guaranteed to be satisfied provided that Eq. (7) holds and $|a_{1}|^{2}+|a_{2}|^{2}+|a_{3}|^{2}+|a_{4}|^{2} \leq 2$. Thus, the constraints can be simplified as
\begin{eqnarray}
|a_{1}|^{2}+|a_{2}|^{2}+|a_{3}|^{2}+|a_{4}|^{2} \leq 1 + |\Delta|^{2} \leq 2.
\end{eqnarray}
Given an operation element $M_{1}$ with its corresponding probability of success $p$, the entries of $M_{1}$ can be multiplied by a complex number $c$ to increase the success probability of the transformation by a factor of $|c|^{2}$. In this case, one has
\begin{eqnarray}
|c|^{2} \leq \frac{|a_{1}|^{2}+|a_{2}|^{2}+|a_{3}|^{2}+|a_{4}|^{2}-\sqrt{(|a_{1}|^{2}+|a_{2}|^{2}+|a_{3}|^{2}+|a_{4}|^{2})^{2}-4|\Delta|^{2}}}{2|\Delta|^{2}}.
\end{eqnarray}
It is obvious from the above expression that the greatest value of $|c|^{2}$, and consequently, of $p|c|^{2}$ will be obtained when $|c|^{2}$ is equal to the right hand side of Eq. (9). As a result, for the transformations having the maximum probability of success, Eq. (8) becomes
\begin{eqnarray}
|a_{1}|^{2}+|a_{2}|^{2}+|a_{3}|^{2}+|a_{4}|^{2} = 1 + |\Delta|^{2}  \leq 2.
\end{eqnarray}
If Eq. (10) is not satisfied for a given operation element, one can easily scale it to give the optimal probability by multiplying with the maximum allowed value of $|c|^{2}$ given by Eq. (9). A special class of transformation schemes, called one successful branch protocols (OSBP), for the distillation of entangled states have been considered in the literature \cite{15,16,17}. This scenario involves $n$ parties performing a unique two outcome POVM, whose operation elements are constructed in a way that after each POVM, one of the two possible resulting states contains no $n$-partite entanglement. For each party, this restriction mathematically implies that $\det[I-M_{1}^{\dag}M_{1}]=0$, assuming the successful branch is realized by the application of $M_{1}$. The fact that this condition is nothing but the equality part of Eq. (10) guarantees the optimality of OSBP in the case of one-shot quantum operations. Finally, it is also possible to show that
the necessary and sufficient conditions given by Eq. (8) are equivalent to the constraint on POVM elements that the eigenvalues of $M_{1}^{\dag}M_{1}$ should be less than or equal to one.

\section{Structure of the FES Subspace}

A simple and systematic method for the classification of FES states has been recently obtained \cite{21}. It has been noted that imposing flip and exchange symmetry on the system drastically simplifies the form of operators. Thus, FES ILOs can be written as
\begin{eqnarray}
M(t)=f(t)
\left(\begin{array}{cc}
1 & t \\
t & 1 \\
\end{array}\right),
\end{eqnarray}
where $t\neq\pm1$. Assuming that $|\psi(0)\rangle$ is a normalized $n$ qubit FES state, all equivalent normalized FES states can then be obtained as
\begin{eqnarray}
|\psi(t)\rangle=\frac{M^{\otimes n}|\psi(0)\rangle}{\sqrt{\langle\psi(0)|(M^{\dagger}M)^{\otimes n}|\psi(0)\rangle}}.
\end{eqnarray}
They lie on a curve parametrized by $t$ provided that $t$ is real. As $t$ changes from $-\infty$ to $\infty$, excluding $t=\pm1$, $|\psi(t)\rangle$ traces the curve. However, if $|\psi(0)\rangle$ turns out to be an eigenstate of $M^{\otimes n}(t)$, no FES ILO will alter it or by definition  $|\psi(0)\rangle$ will form an equivalence class by itself. Eigenstates of $M^{\otimes n}$ are of the form $\otimes^{n}_{k=1}|\pm\rangle_{k}$ where $|\pm\rangle=(1/\sqrt{2})(|0\rangle \pm|1\rangle)$, and number of $|+\rangle$ and $|-\rangle$ states in the Kronecker product are $p$ and $q=n-p$, respectively. Flip symmetric ones are those with even $q$. Eigenvalues are given by
\begin{eqnarray}
\lambda_{pq}=f^{n}(t)(1+t)^{p}(1-t)^{q},
\end{eqnarray}
and they are $n!/p!q!$ fold degenerate. The eigenstate $|\psi_{pq}\rangle$ denotes the FES state obtained by evaluating the symmetric linear combination of degenerate eigenstates corresponding to eigenvalue $\lambda_{pq}$ given by Eq. (13).

In case of three qubits, possible even $q$ values are 0 and 2. While the former corresponds to the separable state $|\psi_{30}\rangle=|+++\rangle$, the latter corresponds to the entangled state $|\psi_{12}\rangle=\frac{1}{\sqrt{3}}(|+--\rangle+|-+-\rangle+|--+\rangle)$, which is equivalent to the $|W\rangle$ state. Since $|GHZ\rangle$ can be written as $|GHZ\rangle=\cos\theta|\psi_{12}\rangle+\sin\theta|\psi_{30}\rangle$ with $\theta=\pi/6$, it lies on the geodesic connecting the separable $|S\rangle$ state and the entangled FES $|W\rangle$ state.
\begin{figure}[H]
\begin{center}
\includegraphics[scale=0.72]{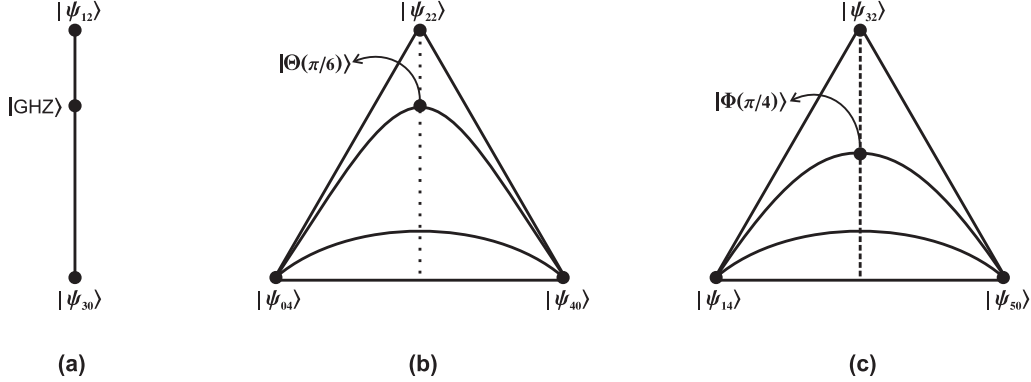}
\caption{Graphical representation of 3, 4 and 5-qubit FES states under ILOs. (a) Almost all states are equivalent to $|GHZ\rangle$ under ILOs while $|W\rangle$ ($|\psi_{12}\rangle$) and $|S\rangle$ ($|\psi_{30}\rangle$) are the neighbors of this
equivalence class. (b) $|\psi_{40}\rangle$ and $|\psi_{04}\rangle$ are the end points of the curves. The dotted line denotes a portion of the great circle $G_{a,a-d,0,d}$ and $|\psi_{22}\rangle$ corresponds to $G_{1,-1,0,2}$. The states lying inside the envelope can be generated using $|\Theta(\theta)\rangle=(\sin\theta/\sqrt{2})(|\psi_{40}\rangle+|\psi_{04}\rangle)+\cos\theta|\psi_{22}\rangle$  with $0 < \theta \leq \pi/2$ as a representative subset. (c)  All curves extend between $|\psi_{14}\rangle$ and $|\psi_{50}\rangle$. The states lying inside the envelope can be generated using the representative subset  $|\Phi(\theta)\rangle=(\sin\theta/\sqrt{2})(|\psi_{50}\rangle+|\psi_{14}\rangle)+\cos\theta|\psi_{32}\rangle$ with $0 < \theta \leq \pi/2$, which is denoted by the dashed line.}
\end{center}
\end{figure}
Allowed $q$ values for four qubits are 0, 2 and 4. The first and the third are separable $|\psi_{40}\rangle$ and $|\psi_{04}\rangle$ states, respectively. The only entangled one is $|\psi_{22}\rangle$ which is nothing but $G_{1,-1,0,2}$ (this state was misprinted as $G_{0,-1,0,1}$ in ref. \cite{21}) in the notation of ref. \cite{7} where $G_{abcd}$ is defined by
\begin{eqnarray}
G_{abcd}=\frac{a+d}{2}(|0000\rangle+|1111\rangle)+\frac{a-d}{2}(|0011\rangle+|1100\rangle)\nonumber \\
+\frac{b+c}{2}(|0101\rangle+|1010\rangle)+\frac{b-c}{2}(|0110\rangle+|1001\rangle).
\end{eqnarray}
Since there are three distinct eigenvalues, the FES subspace is a sphere. All curves start and end on $|\psi_{40}\rangle$ and $|\psi_{04}\rangle$. Expectedly, there exists infinitely many curves corresponding to infinitely many different SLOCC classes. Among the nine classes of four-qubit states, the only FES one is $G_{abcd}$ with $b=d-a$ and $c=0$, and it represents a great circle on the sphere passing through $|\psi_{22}\rangle$ and making equal angles with $|\psi_{40}\rangle$ and $|\psi_{04}\rangle$ \cite{7}. Hence, all four-qubit FES states can be generated, by the application of FES ILOs, using $G_{a,a-d,0,d}$ as a representative subset. If one specifically wants to deal with the curves lying inside the envelope, then considering
$|\Theta(\theta)\rangle=(\sin\theta/\sqrt{2})(|\psi_{40}\rangle+|\psi_{04}\rangle)+\cos\theta|\psi_{22}\rangle$ with $0 < \theta \leq \pi/2$ as a representative is sufficient.

When it comes to five qubits, the only separable eigenstate is $|\psi_{50}\rangle$.
The remaining two are entangled states represented by $|\psi_{32}\rangle$ and $|\psi_{14}\rangle$. The FES subspace is again three dimensional and the curves join
$|\psi_{50}\rangle$ and $|\psi_{14}\rangle$. Since all three distinct eigenstates are perpendicular to each other by construction, all curves lying inside the envelope can be generated by the application of FES ILOs to the representative subset  $|\Phi(\theta)\rangle=(\sin\theta/\sqrt{2})(|\psi_{50}\rangle+|\psi_{14}\rangle)+\cos\theta|\psi_{32}\rangle$ with $0 < \theta \leq \pi/2$.

\section{Optimal Local FES Transformations}

Since FES ILOs given by Eq. (11) have a fixed form, one can scale the operators to obtain the optimal local transformations of multiqubit FES states by multiplying the matrices with the greatest allowed value of the scaling factor $f^{2}(t)$, which is $1/(1+|t|)^{2}$ where $t\in(-1,1)$.

For three qubits, if one assumes that the initial state is $|GHZ\rangle$, then $|\psi(t)\rangle$ tends to the entangled state $|\psi_{12}\rangle$ as $t\rightarrow-1$. However, Fig. 2(a) shows that the maximum probability of obtaining a final state in the close neighborhood of the entangled state $|\psi_{12}\rangle$ decays to zero. On the other hand, $|\psi(t)\rangle$ tends to the separable state $|\psi_{30}\rangle$ as $t\rightarrow1$. It can also be seen from Fig. 2(a) that the probability of obtaining a final state in the vicinity of the separable state $|\psi_{30}\rangle$ is at most $1/4$. Furthermore, the maximum probability of success for transforming an arbitrary initial state $|\Gamma(\theta)\rangle=\cos\theta|\psi_{12}\rangle+\sin\theta|\psi_{30}\rangle$ with $0 < \theta < \pi/2$ to a final state, which is in the vicinity of the separable state $|\psi_{30}\rangle$, is examined. Fig. 2(b) displays that the closer the initial state to the entangled state $|\psi_{12}\rangle$, the more robust it becomes against a possible FES noise source.
\begin{figure}[H]
\begin{center}
\includegraphics[scale=0.73]{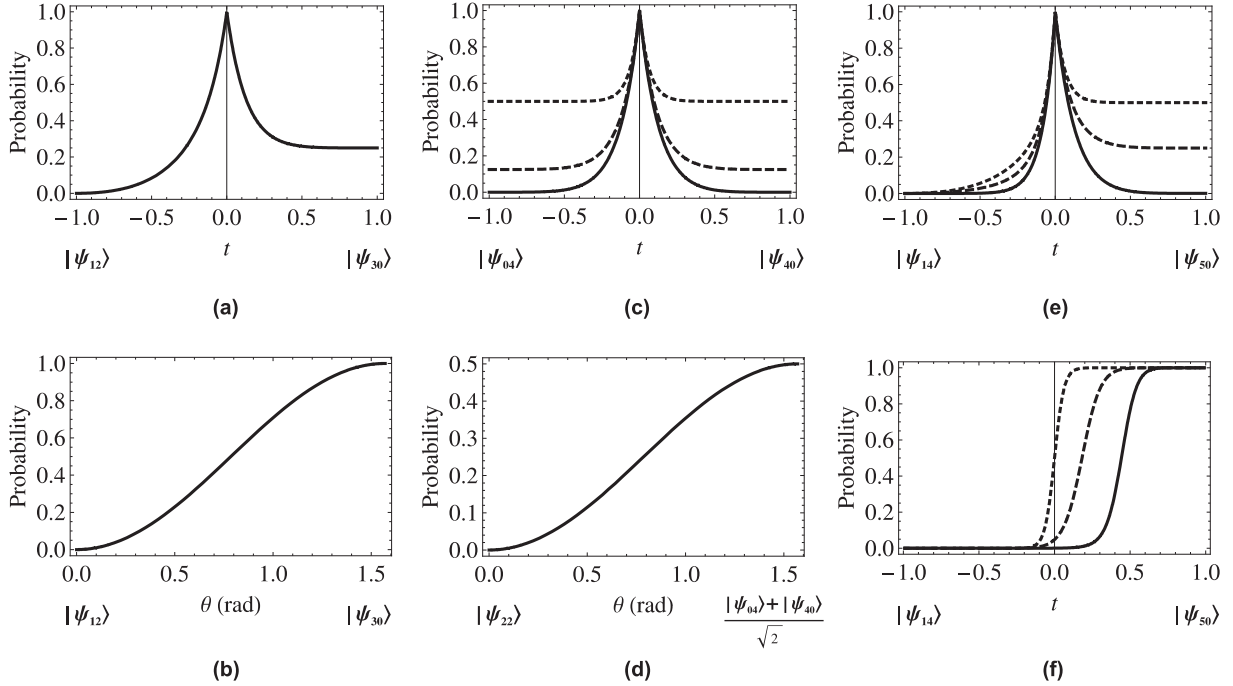}
\caption{Optimal local transformations of FES entangled states. (a) Optimal probabilities of obtaining $|GHZ\rangle$ class FES states starting from the $|GHZ\rangle$ state. (b) Maximum probability of obtaining a final state in the vicinity of $|\psi_{30}\rangle$ assuming that the initial state is $|\Gamma(\theta)\rangle=\cos\theta|\psi_{12}\rangle+\sin\theta|\psi_{30}\rangle$ with $0 < \theta < \pi/2$. (c) Maximum probabilities of obtaining four-qubit FES states, under the assumption that the three initial states are $|\Theta(\pi/100)\rangle$ (solid line), $|\Theta(\pi/6)\rangle=|GHZ_{4}\rangle$ (dashed line) and $|\Theta(\pi/2)\rangle$ (dotted line). (d) Maximum probability of obtaining a final state in the close neighborhood of one of the separable states when the initial state is $|\Theta(\theta)\rangle=(\sin\theta/\sqrt{2})(|\psi_{40}\rangle+|\psi_{04}\rangle)+\cos\theta|\psi_{22}\rangle$ with $0 < \theta < \pi/2$. (e) Maximum probabilities of obtaining five-qubit FES states, under the assumption that the initial states are $|\Phi(\pi/100)\rangle$ (solid line), $|\Phi(\pi/4)\rangle$ (dashed line) and $|\Phi(\pi/2)\rangle$ (dotted line). (f) Maximum probability of obtaining a final state in the vicinity of the separable state $|\psi_{50}\rangle$ when the initial states are arbitrary points on the curves generated from $|\Phi(\pi/2)\rangle$ (dotted line), $|\Phi(\pi/10)\rangle$ (dashed line) and $|\Phi(\pi/100)\rangle$ (solid line).}
\end{center}
\end{figure}
For four qubits, among the infinitely many curves joining $|\psi_{40}\rangle$ and $|\psi_{04}\rangle$, three of them are chosen for the investigation of the optimal FES transformations. As $t\rightarrow1$ ($t\rightarrow-1$), all three initial states get closer and closer to the separable state $|\psi_{40}\rangle$ ($|\psi_{04}\rangle$). Fig. 2(c) shows the probability of success of optimal FES transformations, assuming that the initial states are $|\Theta(\pi/100)\rangle$ (solid line), $|\Theta(\pi/6)\rangle=|GHZ_{4}\rangle$ (dashed line) and $|\Theta(\pi/2)\rangle$ (dotted line). It should be noted that this discussion is fundamentally different from the three-qubit case since the initial states chosen here belong to different FES SLOCC classes. Fig. 2(d) displays the maximum probability of obtaining a final state in the close neighborhood of one of the separable states when the initial state is $|\Theta(\theta)\rangle=(\sin\theta/\sqrt{2})(|\psi_{40}\rangle+|\psi_{04}\rangle)+\cos\theta|\psi_{22}\rangle$ with $0 < \theta < \pi/2$ . Considering the plots, one can conclude that the closer the entangled four-qubit FES states to the entangled state $|\psi_{22}\rangle$, the more robust they become in the sense that the optimum probability of converting them to the states lying in the vicinity of $|\psi_{04}\rangle$ and $|\psi_{04}\rangle$ vanish.

For five qubits, the number of curves, which corresponds to the number of different FES SLOCC classes, are also infinite and again only three of them are considered. While all three initial states tend to the separable state $|\psi_{50}\rangle$ as $t\rightarrow1$, they approach to the entangled state $|\psi_{14}\rangle$ as $t\rightarrow-1$. Fig. 2(e) illustrates the probability of success of optimal FES transformations, assuming that the initial states are $|\Phi(\pi/100)\rangle$ (solid line), $|\Phi(\pi/4)\rangle$ (dashed line) and $|\Phi(\pi/2)\rangle$ (dotted line). The asymmetry in the plot is due to the fact that $|\psi_{14}\rangle$ is an entangled state while $|\psi_{50}\rangle$ is a separable state. Fig. 2(f) shows the maximum probability of obtaining a final state in the close neighborhood of the separable state $|\psi_{50}\rangle$ when the initial states are arbitrary points on the three curves generated from $|\Phi(\pi/2)\rangle$ (dotted line), $|\Phi(\pi/10)\rangle$ (dashed line) and $|\Phi(\pi/100)\rangle$ (solid line). Consequently, five-qubit entangled states, which are in the vicinity of the curve connecting $|\psi_{32}\rangle$ and $|\psi_{14}\rangle$, are more robust than other entangled states.

\section{Conclusion}

In summary, optimal local FES transformations relating two multiqubit FES states have been investigated  by making use of the constraints on the elements of a quantum operation. It has been shown that some entangled FES states are more robust than others under FES ILOs. Namely, for certain initial states, optimal probability of success drops to zero as the final state approaches to a separable state. For example, among the three-qubit $|GHZ\rangle$-equivalent states, those that are closer to the FES $|W\rangle$ state are more robust than the other members of the class. Although our calculations have been limited to the three, four and five-qubit cases, the generalization of the present work to include $n$-qubit FES states is straightforward, since a systematic method has been presented for classifying these states under SLOCC \cite{21}.

\section*{Acknowledgements}

This work has been partially supported by the Scientific and Technological Research Council of Turkey (TUBITAK) under Grant 107T530. The authors would like to thank B. \c{C}akmak for helpful discussions.


\begin{thebibliography}{99}
\bibitem{1}  R. Horodecki, et al., Rev. Mod. Phys. 81 (2009) 865.
\bibitem{2}  C.H. Bennett, D.P. DiVincezo, J.A. Smolin, W.K. Wootters, Phys. Rev. Lett. 76 (1996) 722.
\bibitem{3}  C.H. Bennett, D.P. DiVincezo, J.A. Smolin, W.K. Wootters, Phys. Rev. A 54 (1996) 3824.
\bibitem{4}  C.H. Bennett, S. Popescu, D. Rohrlich, J.A. Smolin, A.V. Thapliyal, Phys. Rev. A 63 (2000) 012307.
\bibitem{5}  W. D\"{u}r, G. Vidal, J.I. Cirac, Phys. Rev. A 62 (2000) 062314.
\bibitem{6}  H.-K. Lo and S. Popescu, Phys. Rev. A 63 (2001) 022301.
\bibitem{7}  F. Verstraete, J. Dehaene, B. De Moor, H. Verschelde, Phys. Rev. A 65 (2002) 052112.
\bibitem{8}  L. Lamata, J. Le\'{o}n, D. Salgado, and E. Solano, Phys. Rev. A 75 (2007) 022318.
\bibitem{9}  L. Borsten, D. Dahanayake, M. J. Duff, A. Marrani, and W. Rubens, Phys. Rev. Lett. 105 (2010) 100507.
\bibitem{10} D. Li, X. Li, H. Huang, and X. Li, Phys. Rev. A 76 (2007) 052311.
\bibitem{11} T. Bastin, S. Krins, P. Mathonet, M. Godefroid, L. Lamata, E. Solano, Phys. Rev. Lett. 103 (2009) 070503.
\bibitem{12} G. Vidal, Phys. Rev. Lett. 83 (1999) 1046.
\bibitem{13} M. A. Nielsen, Phys. Rev. Lett. 83 (1999) 436.
\bibitem{14} B. He, J. A. Bergou, Phys. Rev. A 78 (2008) 062328.
\bibitem{15} A. Ac\'{\i}n, E. Jan\'{e}, W. D\"{u}r, and G. Vidal, Phys. Rev. Lett. 85 (2000) 4811.
\bibitem{16} L.-m Liang, C.-Z Li, Phys. Lett. A 308 (2003) 343.
\bibitem{17} A. Yildiz, Phys. Rev. A 82 (2010) 012317.
\bibitem{18} S. Kinta\c{s}, S. Turgut, J. Math. Phys. 51 (2010) 092202.
\bibitem{19} W. Cui, E. Chitambar, H.-K. Lo, Phys. Rev. A 82 (2010) 062314.
\bibitem{20} W. Cui, W. Helwig, and H. K. Lo, Phys. Rev. A 81 (2010) 012111.
\bibitem{21} Z. Gedik, Opt. Commun. 284 (2011) 681.
\bibitem{22} H.-K. Lo and S. Popescu, Phys. Rev. Lett. 83 (1999) 1459.
\end{thebibliography}
\end{document}